\documentclass[11pt]{article}
\pdfoutput=1
\usepackage{amsmath,amssymb,color,epsfig,cite}
\usepackage{graphicx}
\usepackage{subfigure}
\usepackage{setspace}

\usepackage{amsfonts}

\newcommand{\hoch}[1]{$\, ^{#1}$}

\makeatletter
\@addtoreset{equation}{section}
\makeatother

\newcommand{\be}{\begin{equation}}
\newcommand{\ee}{\end{equation}}
\newcommand{\bea}{\setlength\arraycolsep{2pt} \begin{eqnarray}}
\newcommand{\eea}{\end{eqnarray}}
\newcommand{\nn}{\nonumber}

\def\ft#1#2{{\textstyle{\frac{\scriptstyle #1}{\scriptstyle #2} } }}
\def\fft#1#2{{\frac{#1}{#2}}}

\def\0{{\sst{(0)}}}
\def\1{{\sst{(1)}}}
\def\2{{\sst{(2)}}}
\def\3{{\sst{(3)}}}
\def\4{{\sst{(4)}}}
\def\5{{\sst{(5)}}}
\def\6{{\sst{(6)}}}
\def\7{{\sst{(7)}}}
\def\8{{\sst{(8)}}}
\def\sst#1{{\scriptscriptstyle #1}}

\begin{document}

\begin{center}
{\large {\bf Cosmological Time Crystals From Einstein-Cubic Gravities}}

\vspace{15pt}
{\large Xing-Hui Feng\hoch{1}, Hyat Huang\hoch{2},
Shou-Long Li\hoch{3,4,1*}, H. L\"u\hoch{1*} and Hao Wei\hoch{4}}

\vspace{15pt}

\hoch{1}{\it Center for Joint Quantum Studies and Department of Physics,\\
 School of Science,  Tianjin University, Tianjin 300350, China}

\vspace{10pt}

\hoch{2}{\it College of Physics and Communication Electronics, \\
 Jiangxi Normal University, Nanchang 330022, China}

\vspace{10pt}

\hoch{3}{\it Department of Physics and Synergetic Innovation Center for Quantum Effects and Applications, Hunan Normal University, Changsha, Hunan 410081, China}

\vspace{10pt}

\hoch{4}{\it School of Physics, Beijing Institute of Technology, Beijing 100081, China}

\vspace{40pt}

\underline{ABSTRACT}
\end{center}

By including appropriate Riemann cubic invariants, we find that the dynamics
of classical time crystals can be straightforwardly realized in Einstein gravity on
the FLRW metric. The time reflection symmetry is spontaneously broken
in the two vacua with the same scale factor $a$, but opposite $\dot a$.
The tunneling from one vacuum to the other provides a robust mechanism for bounce universes; it always occurs for systems with positive energy density.  For suitable matter energy-momentum tensor we also construct cyclic universes. Cosmological solutions that resemble the classical time crystals can be constructed in massive gravity.

\vfill
\hoch{*} shoulongli@hunnu.edu.cn\ \ \ mrhonglu@gmail.com

\thispagestyle{empty}

\pagebreak



\newpage

\section{Introduction}

Recently, a fascinating concept of time crystal was proposed \cite{Shapere:2012nq,Wilczek:2012jt}, and subsequently realized in experiments \cite{Zhang1,Choi1,Autti:2017jcw}. A time crystal refers to a ground state that breaks the time-translational invariance, such that it is periodic in both space and time.
The subject has attracted considerable attention, e.g.\cite{Bruno:2013rdc,lgy,Else:2016,yao:2016,
Sacha:2017fqe,yao:2018}. From the view of effective dynamics \cite{Shapere2017}, a minimal classical time crystal can be realized mathematically by including a potential with a well-defined regulator; it is characterized by a non-smooth reversal of the velocity at the boundary. What is unusual is that the sudden velocity reversal is not caused by a ``brick wall'' potential; it is a consequence of spontaneous symmetry breaking analogous to the Higgs mechanism, but in the momentum space.

It is interesting and important to study this idea in the context of gravity. There are many important time-dependent systems in General Relativity (GR) such as the evolution of our Universe, dynamical black holes, gravitational waves and so on. The time crystal behavior of an oscillating scalar field in the expanding Friedmann-Lema\^itre-Robertson-Walker (FLRW) universe was constructed in \cite{Bains:2015gpv, Easson:2016klq,Easson:2018qgr}. The aim of this paper is to study the possibility of treating the universe itself as a time crystal. Such a cosmological model naturally depicts a cyclic universe, which would necessarily violate the null-energy condition (NEC) in GR. Nevertheless, the idea of our Universe being cyclic still attracts much attention, and the most famous model is the ekpyrotic universe, based on string theory \cite{Khoury:2001wf}.  The time crystal mechanism provides an alternative realization.

Constructing cosmological time crystals was attempted in gravity on non-commutativity geometry \cite{Das:2018bzx}. In fact, the dynamics of classical time crystals \cite{Shapere:2012nq,Shapere2017} can be easily realized in gravities, when we include higher-order curvature invariants. In this paper we consider Einstein gravity extended with appropriate Riemann cubic invariants, coupled to some homogeneous and isotropic perfect fluid. A new feature arising is that there is now a Hamiltonian constraint owing to the general diffeomorphism whilst in a classical mechanical system, the Hamiltonian yields an arbitrary conserved energy that should be bounded below.

In section 2, we construct the cubic invariants that facilitate the time crystal mechanism of \cite{Shapere:2012nq,Shapere2017}. We study the properties of the resulting cosmological solutions. In section 3, we present three explicit examples based on different types of the perfect fluid models. In the last example, we consider massive gravity instead of Einstein gravity.
We discuss and conclude the paper in section 4.

\section{The theory and the cosmological model}

A crucial ingredient in our construction involves the Riemann cubic invariants, which have in general eight terms:
\bea
{\cal L}_{\3} &=& \sqrt{-g}\Big(e_1 R^3 + e_2 R\,R_{\mu\nu} R^{\mu\nu} + e_3 R^{\mu}_{\nu} R^{\nu}_\rho R^{\rho}_\mu + e_4 R^{\mu\nu} R^{\rho\sigma} R_{\mu\rho\nu\sigma}+ e_5 R R^{\mu\nu\rho\sigma} R_{\mu\nu\rho\sigma}\cr
&&+e_6 R^{\mu\nu} R_{\mu \alpha\beta\gamma} R_{\nu}{}^{\alpha\beta\gamma} +
e_7 R^{\mu\nu}{}_{\rho\sigma} R^{\rho\sigma}{}_{\alpha\beta} R^{\alpha\beta}{}_{\mu\nu}+e_8 R^\mu{}_\nu{}^\alpha{}_\beta R^\nu{}_\rho{}^\beta{}_\gamma R^{\rho}{}_\mu{}^\gamma{}_{\alpha}
\Big)\,.
\eea
In de Sitter (dS) or anti-de Sitter (AdS) spacetimes, these generate additional linear massive scalar and spin-2 modes.  Decoupling of the ghost-like spin-2 mode requires (e.g.~\cite{Sisman:2011gz})
\be
12 e_2+9 e_3+5 e_4+48 e_5+16 e_6+24 e_7-3 e_8=0\,.
\ee
In this paper, we study cosmology in the FLRW metric
\be
ds^2= -dt^2 + a(t)^2 (dx_1^2 + dx_2^2 + dx_3^2)\,.\label{flrw-flat}
\ee
In the effective Lagrangian, the absence of both $\ddot a^2$ and $\ddot a^3$ terms requires
\bea
216 e_1+60 e_2+18 e_3+16 e_4+48 e_5+14 e_6+12 e_7+3 e_8 &=& 0\,,\nn\\
36 e_1+12 e_2+5 e_3+3 e_4+12 e_5+4 e_6+4 e_7&=&0\,.
\eea
The first equation decouples the massive scalar mode. Intriguingly these are precisely the same conditions for the holographic $a$-theorem \cite{Li:2017txk}. For the linear $\ddot a$ term, we perform integration by parts so that the Lagrangian involves only the $\dot a$ term. With these conditions, we find that the effective Lagrangian is given by
\be
L_{\3} = \ft25\lambda \fft{\dot a^6}{a^3}\,,\qquad
\lambda=2(36 e_1 + 6 e_2 + e_4 + 4 e_5)\,.\label{effectivelag}
\ee
The reason we end up with four parameters is that the cubic Euler density combination vanishes in four dimensions. We may restrict further to Ricci polynomials, namely $7R^3 -36 R\,R_{\mu\nu} R^{\mu\nu} + 36 R^{\mu}_{\nu} R^{\nu}_\rho R^{\rho}_\mu$ \cite{Li:2017ncu}. The analogous construction for quadratic invariants yields the Weyl-squared term, which gives no contribution to equations of motion for the FLRW metric.

We focus on Einstein gravity coupled to some perfect fluid with energy-momentum tensor $({{T^{a}}_b})_m = {\rm diag}(\rho_m,p_m,p_m,p_m)$, and the Riemann cubics.  The Einstein equations yield
\be
\fft{3\dot a^2}{a^2} - \lambda \fft{ \dot a^6}{a^6}= \rho_m\,,\qquad
-\fft{2\ddot a}{a} - \fft{\dot a^2}{a^2}  + \lambda \Big(\fft{2\dot a^4 \ddot a}{a^5}-
\fft{\dot a^6}{a^6}\Big)= p_m\,.\label{gencosmoeq}
\ee
For simplicity, we consider an effective theory where the energy density is a function of the scale factor $a$.  For example, the vacuum energy density is given by $\rho_m=\Lambda_0$, the bare cosmological constant. In radiation or matter dominated universes, we have $\rho_m\sim a^{-4}$ and $a^{-3}$ respectively. For a free massless scalar $\phi$, we have instead $\rho_m\sim a^{-6}$.  If we express $\rho_m=V(a)/(2a^3)$, the energy-momentum conservation requires $p_m= -V'(a)/(6a^2)$, giving rise to the equation of state $w=-{a V'}/(3V)$.

The equations of motion can now be derived from the effective Lagrangian
\be
L = -6 a \dot a^2 + \ft25\lambda\, \fft{\dot a^6}{a^3} - V\,.\label{efflag}
\ee
The negative kinetic energy $-\dot y^2/2$ proposed in \cite{Shapere:2012nq,Shapere2017} is hard to justify in classical mechanics, it arises naturally in gravity.  This term should not be viewed as ghost-like owing to the general diffeomorphism, which imposes the Hamiltonian constraint
\be
H=-6 a \dot a^2 + 2\lambda\, \fft{\dot a^6}{a^3} + V=0\,.\label{Hconstraint}
\ee
In fact this constraint is equivalent to the first equation in (\ref{gencosmoeq}).  By contrast, although $H$ is conserved in classical mechanics, it does not necessarily vanish.

\begin{figure}[ht]
\ \ \ \ \ \includegraphics[width=6cm]{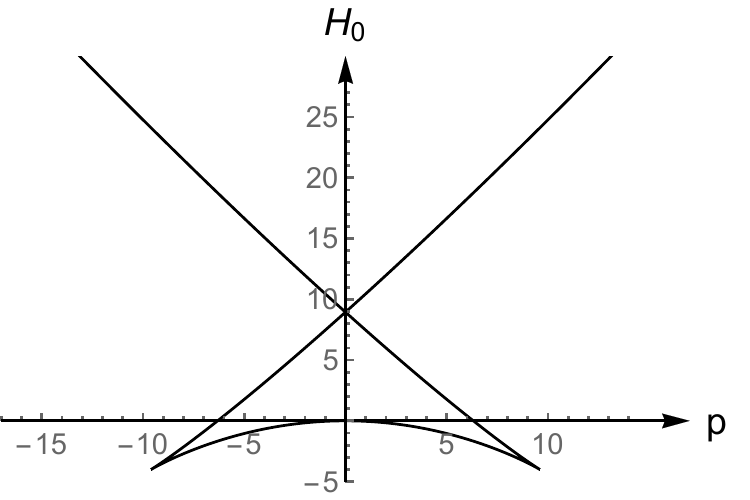}
\ \ \ \ \ \includegraphics[width=6cm]{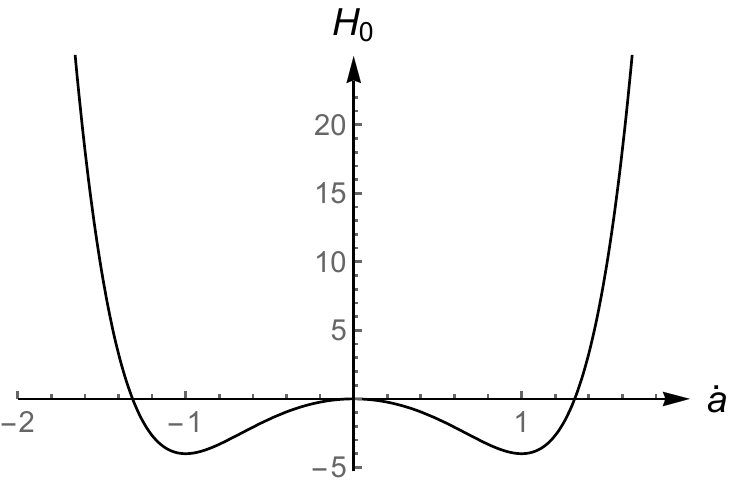}
\caption{\small The sparrow tail in the left plot is characteristic when we include a higher-order kinetic term. The cusp singularities imply that $\dot a$ cannot be continuous at the tip. Instead the system tunnels from one vacuum to the other, keeping $\dot a^2$ continuous. The right plot shows that $H_0(\dot a)$ shapes like a two-dimensional Mexican hat. In both plots we set $\lambda=1$ and $a=1$. For general quantities, the true vacua are at $\dot a=\pm \lambda^{1/4} \,a$, corresponding to $H_0^{\rm min}=-4a^3/\sqrt{\lambda}$.}
\label{H0shape}
\end{figure}

In this paper, we shall consider only $\lambda>0$, such that the gravitational part of the Hamiltonian, $H_0=-6 a \dot a^2 + 2\lambda\, {\dot a^6}/{a^3}$, is bounded below. A key property is that in terms of the canonical momentum
\be
p=\fft{\partial L}{\partial \dot a}=-12 a \dot a \big(1 - \fft{\lambda\dot a^4}{5a^4}\big)\,,
\ee
$H_0(p,a)$ can be multi-valued and the minimum of $H_0$ occurs not when $p=0$, but when
\be
\fft{\partial H_0}{\partial \dot a}=0\,,\qquad \Longrightarrow\qquad
\dot a = \pm \lambda^{-\fft14} a\,,\qquad H_{0}^{\rm min}=-\fft{4a^3}{\sqrt{\lambda}}\,.\label{turning1}
\ee
The $p=0=\dot a$ point is instead a local maximum. This is analogous to the Higgs mechanism, but in the momentum space.  $H_0$ as function of $p$ is depicted in Fig.~\ref{H0shape}.

The evolution of $a(t)$ near $H_0^{\rm min}$ depends on the potential $V(a)$. In order to study this behavior, we also examine $H_0$ as a function of $\dot a$, depicted also in Fig.~\ref{H0shape}.  The Hamiltonian constraint (\ref{Hconstraint}) at $H_0^{\rm min}$ implies that
\be
\widetilde V(a)\equiv V(a) - \fft{4a^3}{\sqrt{\lambda}}=0\,.\label{turning2}
\ee
If it has no solution, (e.g.~$V=-3a^3$, for a negative cosmological constant,) then $H_0^{\rm min}$ can never be reached.  If equation (\ref{turning2}) has a solution at $a=a_0$, the system may reach $a_0$, but cannot stay there since $\dot a\ne 0$. Thus we must require that
the cubic equation (\ref{Hconstraint}) for $z=\dot a^2$ has a positive root, and it does if and only if $\widetilde V(a)\le 0$ in the connected region. In fact, there are two positive roots, given by
\be
\dot a^2 = \fft{2a^2}{\sqrt{\lambda}} \cos\fft13\Big(
\arccos\big[\fft{a_0^3}{a^3} \fft{V(a)}{V(a_0)}\big]-k\pi\Big)\,,\qquad k=\pm 1.
\ee
The third root, corresponding to $k=3$, is negative and should be ignored.
In the vicinity of $a=a_0$, we have
\be
\dot a^2 = \fft{a_0^2}{\sqrt{\lambda}} \Big(1+k \sqrt{\ft{2}{a_0}\big(1+w(a_0)\big)(a-a_0)}
+ \cdots\Big)\,.
\ee
For matter satisfying the NEC ($w\ge -1$), we must have $a\ge a_0$, thus a bounce occurs at $a=a_0$. Furthermore, at $a=a_0$, we have two vacua with $\dot a_0^\pm=\pm \lambda^{-1/4} a_0$ respectively, and they are energy degenerate, but break the time reflection symmetry.  The tunneling from one vacuum to the other keeps $\dot a^2$ continuous, but causes $\dot a$ to jump from the negative to the positive, or vice versa, analogous to the situation when a pingpong hits a brick wall.

It is instructive to determine whether there should be an external ``brick wall'' source, since $\ddot a$, and hence the curvature, has a $\delta$-function singularity of the comoving time. Assuming that the turning point $a=a_0$ occurs at $t=0$, we can solve the function $a(t)$ at small $t$. The solution is smooth except at $t=0$, and the extra source required for the bouncing behavior is formally given by
\be
\rho_{\rm ext}=0\,,\qquad
p_{\rm ext}=4\sqrt{2} k a_0^{-1/2} \lambda^{-3/8}\sqrt{ 1+w(a_0)}\, \sqrt{|t|}\, \delta (t)\,, \qquad k=-1,1\,.
\ee
Since the energy-momentum conservation does not involve a time derivative of $p$, we can effectively treat $p_{\rm ext}=0$.  It was demonstrated \cite{Shapere2017} that this matter source can be replaced by some well-defined regulator in classical time crystals, in which case, it is of great interest to analyse the energy condition of the external source.

The physical picture is clear.  Owing to the spontaneous symmetry breaking, the cosmology splits into two energy-degenerated vacua, with $\dot a_0^\pm$ respectively. As the universe shrinks to $a_0$, it tunnels from the $\dot a_0^-$ vacuum to the $\dot a_0^+$ one and starts to expand, creating a bounce at $a=a_0$.  It follows from (\ref{turning2}) that this bounce mechanism is robust and will always occur for positive energy density.

\section{Explicit examples}

\begin{figure}[ht]
\centerline{
\includegraphics[width=6cm]{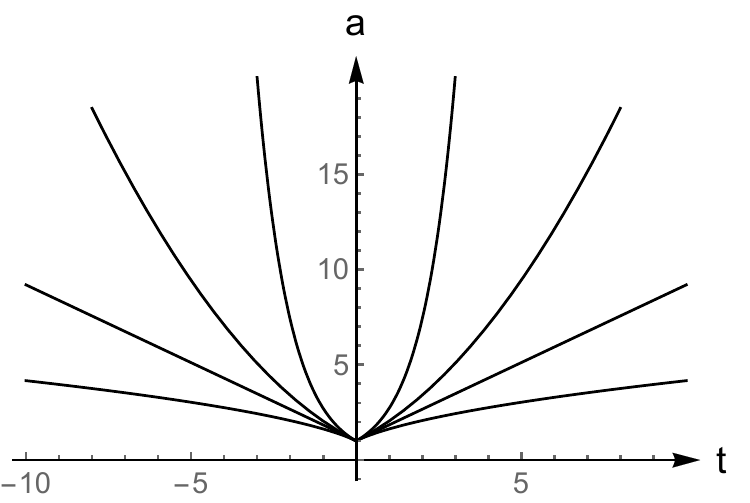}\ \ \ \ \
\includegraphics[width=6cm]{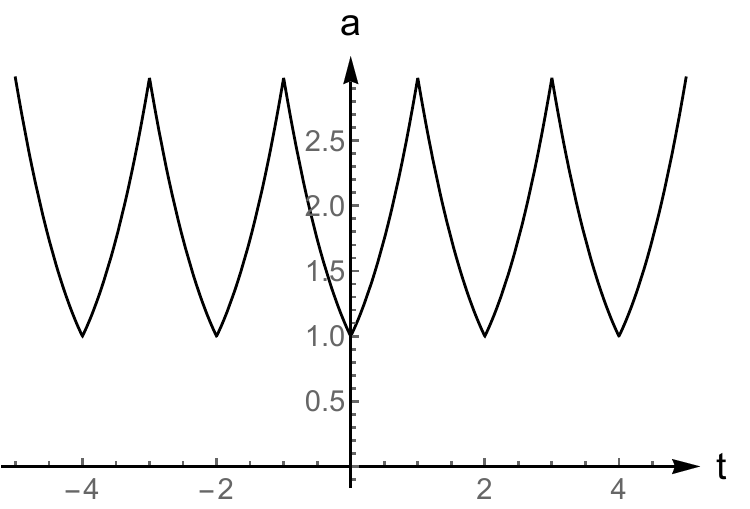}}
\caption{\small The left plot shows bounce universes, where $\dot a$ reverses its sign at $t=0$, $a_0=1$. We have chosen $\lambda=1$ and $w=-1,-2/3, -1/3, 1/3$ for the lines from the top to the bottom.  The right plot is the cosmological time crystal where $\dot a$ reverses signs at both $a_\pm=1,3$. The theory involves massive gravity, discussed in the third example.}
\label{bounce}
\end{figure}

\noindent{\bf I. Bounce universes}: The simplest example is perhaps when $w$ is a constant, for which $V=2q^2 a^{-3w}$ where $q$ is a constant. For $\lambda=0$, the universe is expanding with $a=(\fft34(1+w)^2q^2 t^2)^{{1}/({3(1+w)})}$, with an initial spacetime singularity at $t=0$.  For non-vanishing $\lambda$, a bounce must occur, at $a_0^{3(1+w)}=\sqrt{\lambda} q^2/2$.
The $k=-1$ solution is governed by
\be
\dot a^2=\fft{2}{\sqrt{\lambda}} a^2 \cos\ft13\Big(\arccos\big[(\fft{a_0}{a})^{3(1+w)}
\big]+\pi\Big)\,.
\ee
Thus we see that for the standard cosmology with positive energy density, the introduction the Riemann cubics generates a bounce universe. In Fig.~\ref{bounce}, we plot the $a(t)$ solutions for various $w$, taking $a_0=1$. It is worth pointing out that our numerical analysis indicates that the solutions are stable against small perturbations of the initial condition $a(0)=1$. Furthermore, the bounce scenario should be distinguished from those in literature where $\dot a=0$ at the time of bounce.  The consequence is that the NEC must be violated in the framework of Einstein gravity.  In our case, $\dot a$ does not vanish at the time of bounce, but tunnels from the negative value to the positive value in such a way that the matter system satisfies the NEC.

\begin{figure}[ht]
\centerline{
\ \ \ \ \ \includegraphics[width=6cm]{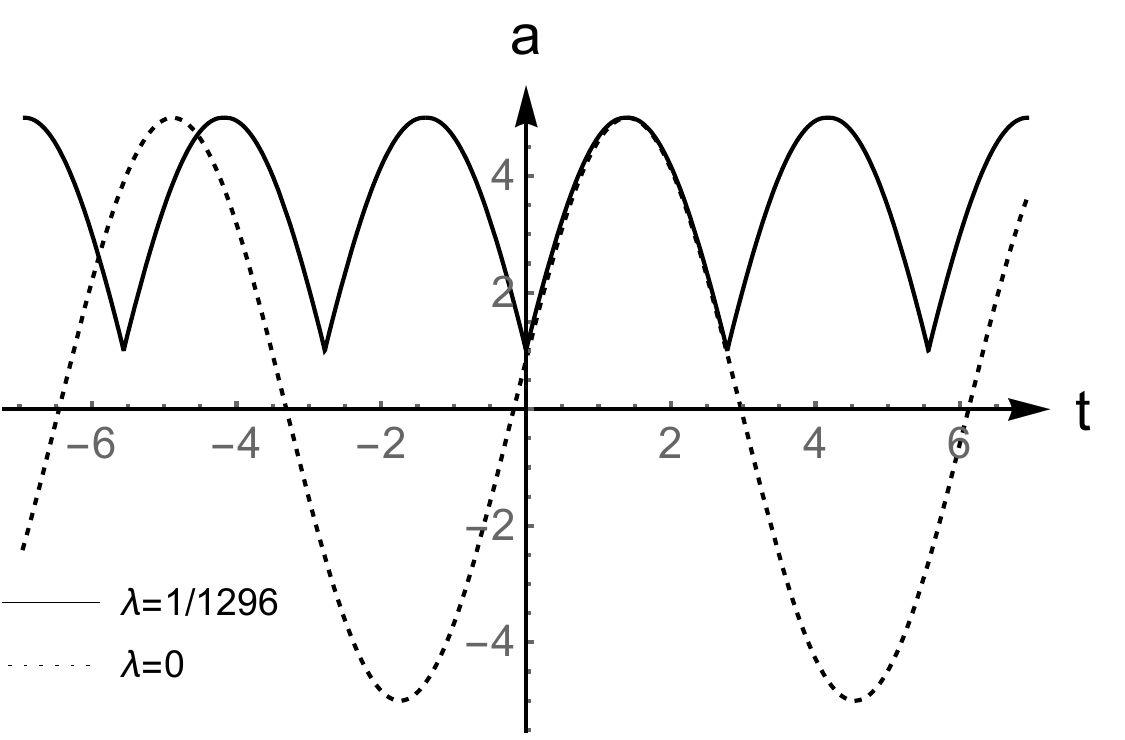}\ \ \ \
\includegraphics[width=6cm]{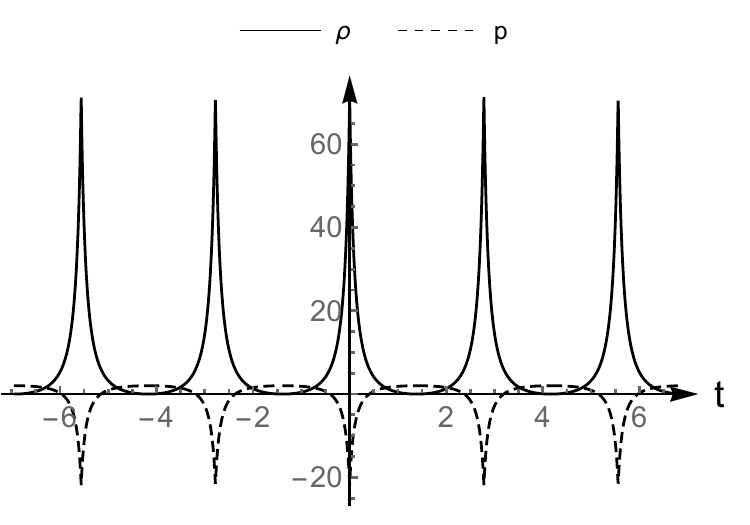}}
\caption{\small A cyclic universe due to the time-crystal mechanism is depicted by the solid line of the left plot ($a_0=1$, $A=5$, $\ell=1$, $k=-1$, corresponding to $\Lambda=-3$ and $\alpha=5$.) The dotted sinuous line is an exact solution of $a$ when $\lambda=0$.  In the right $\rho$ and $p$ of the solid $a$ solution are presented and it is easy to see that $\rho+p >0$.}
\label{cyclic}
\end{figure}

\noindent{\bf II. Cosmological time crystals}: If the potential $V$ has a zero at finite $a=A>a_0$; furthermore, $\widetilde V(a)<0$ for $a\in (a_0, A)$, then $a(t)$ shrinks smoothly from $a=A$, until at $a=a_0$ where it bounces, creating a cyclic universe. As a concrete example, we consider $V=2\Lambda_0 a^3 + 6\alpha^2 a$, where the second term can be generated by a sigma model \cite{Geng:2014vza}.
For negative cosmological constant $\Lambda_0=-3/\ell^2$, the potential vanishes at $A=\alpha \ell$.  In fact, for $\lambda=0$, an exact solution can be found, namely $
a=A \sin(t/\ell)$.  The solution appears to be cyclic, but the corresponding universe is not owing to the curvature singularity at $a=0$.  When $\lambda$ is included, the $\dot a^6$ term has no effect at $a=A$, where $\dot a=0$.  However, there is a turning point $a_0$:
\be
0<a_0= \fft{\sqrt3\,\alpha\ell \lambda^{1/4}}{\sqrt{2\ell^2 + 3 \sqrt{\lambda}}}<A\,,
\ee
where the universe bounces. The solution is depicted in Fig.~\ref{cyclic}.  It should be pointed out that for this particular model, the time crystal mechanism is only possible for the negative cosmological constant, since $\alpha^2$ must be positive for the sigma model with the standard kinetic term \cite{Geng:2014vza}. Cosmological time crystals involving a positive cosmological constant will be presented next.

\noindent{\bf III. Cosmological time crystals from massive gravity}: In \cite{Shapere2017}, time crystals typically have two jumping points.  In our gravity model, for $V(a)$ satisfying NEC, there can only be one, causing bounce of the universe. In order to reproduce the analogous behaviors of \cite{Shapere2017}, we have to consider theory beyond Einstein.  We find such a solution exists in dRGT (de~Rham-Gabadadze-Tolley) massive gravity of \cite{deRham:2010kj} together with the Riemann cubics. The idea of describing the massive graviton was first proposed by Fierz-Pauli~\cite{Fierz:1939ix} in 1939. However, due to the Boulware-Deser (BD) ghost~\cite{Boulware:1973my} of the interactions for massive spin-2 fields in the Fierz-Pauli theory, few important developments had been made in the past decades, until the dRGT theory came along, which is a ghostfree realization of massive gravity and has attracted great attention in the General Relativity community. We refer to e.g.~\cite{Hinterbichler:2011tt, deRham:2014zqa} and references therein for a comprehensive introduction of massive gravity. The Lagrangian of dRGT gravity together with the Riemann cubics is given by:
\bea
{\cal L} &=& \sqrt{-g} \Big(R - 2\Lambda_0 + m^2 ( {\cal U}_2 +c_3 {\cal U}_3 +c_4 {\cal U}_4 )\Big) +\lambda {\cal L}_\3\,,\nn\\
{\cal U}_2 &=& [{\cal K}]^2 -[{\cal K}^2]\,, \qquad
{\cal U}_3 = [{\cal K}]^3 -3 [{\cal K}] [{\cal K}^2] +2 [{\cal K}^3] \,,\nn\\
{\cal U}_4 &=& [{\cal K}]^4 -6 [{\cal K}^2] [{\cal K}]^2 +
8 [{\cal K}^3] [{\cal K}] +3[{\cal K}^2]^2 -6[{\cal K}^4]\,,\nn\\
{{\cal K}^\mu}_\nu &=& {\delta^\mu}_\nu -\sqrt{g^{\mu\lambda} \partial_\lambda \phi^a \partial_\nu \phi^b \eta_{a b}}\,,\qquad \eta_{a b} = \textup{diag}(-1,1,1,1)\,.
\eea
${\cal U}_i$ are interaction potentials, and the rectangular brackets denote traces, e.g.~$[{\cal K}] = \textup{Tr}({\cal K}) ={{\cal K}^\mu}_\mu$.  For the  corresponding St\"uckelberg fields $\phi^a = a_m (0, x_1, x_2, x_3)$, which was introduced to restore the diffeomorphism invariance~\cite{ArkaniHamed:2002sp, deRham:2014zqa}, the effective Lagrangian for the FLRW metric is given by (\ref{efflag}) with
\be
V= 2 \Lambda_0 a^3 +6 m^2(a_m -a ) \Big((4 c_3+4 c_4+2)a^2-a_m (5 c_3 +8 c_4 +1) a +a_m^2 (c_3+4 c_4)\Big)\,,
\ee
where $a_m >0$. For a concrete demonstration, we choose
\be
c_3=\ft2{87}\,,\qquad c_4=\ft7{348}\,,\qquad
\Lambda_0=\ft{5989}{2900}\,,\qquad\lambda=1\,,\qquad m=\ft1{10}\,,\qquad a_m=10\,.
\ee
(Note that in this case, the cosmological constant $\Lambda_0$ is positive here.) The system has two turning points $(a_-,a_+)=(1,3)$. The cosmology cycles between $a_\pm$, as shown in the right plot of Fig.~\ref{bounce}. Since we have $a_\pm <a_m$ in this solution, there is no ghost excitation from the massive gravity sector \cite{DeFelice:2012mx}.

\section{Conclusions and discussions}

Time crystals can arise when the time translational symmetry in the vacuum is spontaneously broken. A simple classical mathematical model \cite{Shapere:2012nq,Shapere2017} involves a ghost-like kinetic $-\dot y^2/2$ augmented by the higher-order $\dot y^4/12$, such that the true vacuum is shifted down with non-vanishing velocity in a specific direction, hence breaking the time reflection symmetry. Such a system is hard to realize in classical mechanics, but it arises naturally in the effective Lagrangian in Einstein gravity on the FLRW metric, extended with appropriate higher-order curvature invariants.

We focused on a class of Riemann cubics and found that for matter satisfying NEC, the time crystal mechanism could generate bounces. The sudden change of the sign of $\dot a$ at the bounce is the effect of tunneling from one vacuum to the other while keeping $\dot a^2$ continuous.  Our analysis shows that this is a robust mechanism for bounce universes; it always occurs for positive energy density. Cyclic universes can also be constructed since shrinking $a(t)$ from its maximum is consistent with NEC.  We also considered massive gravity and constructed time crystals with two sudden reversing points.

Although we have restricted the Riemann cubics such that they do not generate linear ghosts in maximally-symmetric spacetimes, it is import to verify whether they may generate ghosts in our cosmological crystals.  The general perturbation $g_{\mu\nu}= \bar{g}_{\mu\nu} +h_{\mu\nu}$ for high-order gravities can be enormously complicated.  We consider here the scalar perturbations ${h_\mu}^{\nu}$, which in Newtonian gauge is given by
\be
{h_\mu}^{\nu} = \textup{diag} (-2 \Psi, 2 \Phi, 2 \Phi, 2 \Phi) \,.
\ee
The gravitational part of the linear equations of motion in the FLRW metric is given by
\begin{align}
\begin{split}
\delta E_{tt} &\equiv (\frac{6 \dot{a}}{a}-\frac{6 \lambda  \dot{a}^5}{a^5}) \dot{\Phi} +(\frac{2 \kappa ^4 (2 e_2+3 e_3-8 e_5) \left(\dot{a}^2-a \ddot{a}\right)}{a^6}-\frac{4 \lambda  \dot{a}^6}{a^6}) \Psi + (\frac{2 \kappa ^2}{a^2} \\
&\quad -\frac{\lambda (a \kappa ^4 \ddot{a} -\kappa ^4 \dot{a}^2 +2 \kappa^2 \dot{a}^4)}{a^6}) \Phi  \,,
\end{split} \\
\begin{split}
\delta {E^\mu}_{\mu} &\equiv \frac{6 \left(\lambda  \dot{a}^4-a^4\right)}{a^4} \ddot{\Phi} +\frac{24 \left(\lambda  \dot{a}^3 \ddot{a}-a^3 \dot{a}\right)}{a^4} \dot{\Phi} +\frac{6 \left(\lambda  \dot{a}^5-a^4 \dot{a}\right)}{a^5} \dot{\Psi} +\Big(\frac{\kappa ^2 \lambda  \dot{a}^2 \left(\kappa ^2-4 \dot{a}^2\right)}{a^6} \\
&\quad -\frac{2 \left(2 a^4 \kappa ^2+\left(2 e_2+3 e_3-8 e_5\right) \kappa ^4 \dot{a}^2\right)}{a^6}+\frac{\ddot{a}}{a^5} ({2 \left(2 e_2+3 e_3-8 e_5\right) \kappa ^4} \\
&\quad -{\lambda  \left(\kappa ^4-8 \kappa ^2 \dot{a}^2\right)}) \Big) \Phi + \Big(\frac{\ddot{a}}{a^5} ( {\lambda  \left(36 \dot{a}^4 -\kappa ^4\right)} +{2 \left(\left(2 e_2+3 e_3-8 e_5\right) \kappa ^4-6 a^4\right)}) \\
&\quad  +\frac{\lambda  \dot{a}^2 (\kappa ^4-2 \kappa ^2 \dot{a}^2-12 \dot{a}^4)}{a^6}   -\frac{2 \left(\left(2 e_2+3 e_3-8 e_5\right) \kappa ^4 \dot{a}^2 -a^4 \left(\kappa ^2-6 \dot{a}^2\right)\right)}{a^6} \Big) \Psi \,.
\end{split}
\end{align}
where $\kappa$ is defined by $\vec \nabla^2 \Psi = -\kappa^2 \Psi$ and $\lambda$ is given by (\ref{effectivelag}). We thus see that there is no higher-order time derivatives on $\Psi$ in the equations of motion. This implies that the scalar perturbation of our cosmological time crystals can be ghost free. In four dimensions, it is difficult to decouple the ghost freedoms in higher order gravities. The absence of the ghosts in the scalar perturbation in the FLRW model makes our cubic gravities interesting candidate for studying cosmology. There are however many questions that remain. For these higher-derivative time crystal models to be viable, it is necessary to investigate the unitary also for all the vector and tensor modes. Furthermore, it would be interesting to study whether there are ranges of model parameters such that the expanding phase occurs on the cosmological timescales of our universe. While the generalizations to higher dimensions and/or involving higher-order polynomial curvature invariants are straightforward, it is nevertheless of interest to work the detail to see whether new phenomena could arise.

\section*{Acknowledgement }

We are grateful to Yi-Fu Cai, Yue-Zhou Li, Jiro Soda, Zhao-Long Wang and Zhang-Qi Yin for useful discussions. XHF, HH and HL are supported in part by NSFC grants No.~11875200 and No.~11935009. SLL and HW are  supported in part by NSFC grants No.~11975046, No.~11575022, No.~11175016, No.~11947216,  and China Postdoctoral Science Foundation 2019M662785.

\end{document}